\documentclass[aps,prl,twocolumn,superscriptaddress]{revtex4}
\usepackage[intlimits]{amsmath}
\usepackage{amsfonts}
\usepackage{graphics}
\usepackage{subfigure}
\usepackage[usenames]{color}
\usepackage{epstopdf,epsfig}
\usepackage{xcolor}
\usepackage{indentfirst}
\usepackage{enumerate}

\usepackage{graphicx}
\usepackage{amsmath}
\usepackage{amssymb, mathtools, mathrsfs}
\usepackage{hyperref}

\newcommand\be            {\begin{equation}}
\newcommand\ee            {\end{equation}}

\def\bea{\begin{eqnarray}}
\def\eea{\end{eqnarray}}

\newcommand\EN           {\end{equation}}
\newcommand\bes           {\begin{subequations}}
\newcommand\esu           {\end{subequations}}

\def\3pt#1#2#3{{\langle{#1}\vert{#2}\vert{#3}\rangle}}

\begin{document}

\title{Entanglement of two disjoint intervals in conformal field theory \\ 
and the 2D Coulomb gas on a lattice}

\author{Tamara Grava}
\affiliation{SISSA, via Bonomea 265, 34136, Trieste, Italy}
\affiliation{School of Mathematics, University of Bristol, Fry Building, Bristol,
BS8 1UG, UK}
\author{Andrew P. Kels}
\affiliation{SISSA, via Bonomea 265, 34136, Trieste, Italy}
\author{Erik Tonni}
\affiliation{SISSA, via Bonomea 265, 34136, Trieste, Italy}
\affiliation{INFN, Sezione di Trieste, via Valerio 2, 34127, Trieste, Italy}

\begin{abstract} 

\noindent
In the conformal field theories given by the Ising and Dirac models,
when the system is in the ground state,
the moments of the reduced density matrix of two disjoint intervals and of its partial transpose
have been written as
partition functions on higher genus Riemann surfaces with $\mathbb{Z}_n$ symmetry.
We show that these partition functions can be expressed as 
the grand canonical partition functions of the 
two-dimensional two component classical Coulomb gas 
on certain circular lattices at specific values of the coupling constant.
\end{abstract}

\maketitle

\textit{Introduction.---}Entanglement in $1+1$ dimensional conformal field theories (CFT)
has attracted a lot of interest during the last two decades,
allowing to explore many-body quantum systems at criticality
and quantum gravity through the AdS/CFT correspondence
(see the reviews in Ref.~\cite{ent-reviews-09}).
The classical Coulomb gas in two spatial dimensions (2D)
occurs in many interesting models of statistical physics
\cite{zinn-justin,kananoff,coulomb-gas-reviews,samaj-book}.
In this Letter we find that some entanglement quantifiers 
in two $1+1$ dimensional CFTs
can be interpreted in terms of a classical 2D Coulomb gas on 
specific circular lattices.

Consider a spatial bipartition $A \cup B$ of a quantum system
whose Hilbert space is factorized as 
$\mathcal{H} = \mathcal{H}_A \otimes \mathcal{H}_B$.
When the entire system is in a pure state $| \Psi \rangle$
(e.g., the ground state), 
the bipartite entanglement is measured by the entanglement entropy 
$S_A = - \,\textrm{Tr}( \rho_A \log \rho_A)$, 
i.e., the von Neumann entropy of the reduced density matrix 
$\rho_A = \textrm{Tr}_{\mathcal{H}_B} | \Psi \rangle \langle \Psi |$
of the subsystem $A$
(normalized by $\textrm{Tr}_{\mathcal{H}_A} \rho_A =1$).
The entanglement entropy can be obtained through the replica limit 
\cite{ent-replica-cft}
\be
\label{ee-replica-def}
S_A =  - \, \partial_n \textrm{Tr} \rho_A^n \big|_{n=1}\,,
\ee
which requires the analytical continuation in $n$ 
of the moments $\textrm{Tr}\rho_A^n$ of $\rho_A$, 
defined for integer $n \geqslant 2$.
The replica limit (\ref{ee-replica-def}) can be written also in terms of 
the R\'enyi entropies $S_A^{(n)} = \tfrac{1}{1-n}  \log \textrm{Tr}\rho_A^n$
as $ S_A = \lim_{n \to 1} S_A^{(n)} $.

When the subsystem $A = A_1 \cup A_2$ 
is the union of two disjoint regions $A_1$ and $A_2$,
its reduced density matrix $\rho_A$ describes a mixed state
whose bipartite entanglement can be evaluated by the  logarithmic negativity 
$\mathcal{E} = \log \textrm{Tr} | \rho_A^{T_2}|$ \cite{neg-def}.
This entanglement quantifier requires us to evaluate 
the trace norm of the partial transpose $\rho_A^{T_2}$,
normalized by  $\textrm{Tr} \rho_A^{T_2} = 1$,
whose matrix elements are defined as
$\langle e_i^{(1)} e_j^{(2)} | \rho_A^{T_2}| e_k^{(1)} e_l^{(2)}\rangle 
= \langle e_i^{(1)} e_l^{(2)} | \rho_A| e_k^{(1)} e_j^{(2)}\rangle $,
with $| e_i^{(1)} \rangle $ and $| e_i^{(2)} \rangle $ being
bases for the Hilbert spaces $\mathcal{H}_{A_1}$ and $\mathcal{H}_{A_2}$
respectively.
The logarithmic negativity $\mathcal{E}$ can be found also
through the following replica limit \cite{cct-neg}
\be
\label{neg-replica-def}
\mathcal{E} \,= 
\lim_{n_e \to 1}
\log  \textrm{Tr} \big(\rho_A^{T_2}\big)^{n_e}\,,
\ee
where the analytic continuation involves 
the moments $\textrm{Tr} \big(\rho_A^{T_2}\big)^{n}$
having even $n=n_e$.


In the case of
a $1+1$ dimensional CFT on the line and in its ground state,
the moments $\textrm{Tr} \rho_A^n$ and $\textrm{Tr} \big(\rho_A^{T_2}\big)^{n}$ 
with integer $n \geqslant 2$
are insightful entanglement quantifiers because 
they encode all the CFT data of the model
\cite{EE-2int-initial, cct-09, cct-11, cct-neg}.
They can be obtained as the partition functions of the CFT 
on genus $g=n-1$ Riemann surfaces with $\mathbb{Z}_n$ symmetry
\cite{cct-09, cct-11, cct-neg,Enolski-Grava-03}.
Analytic expressions for $\textrm{Tr} \rho_A^n$ and $\textrm{Tr} \big(\rho_A^{T_2}\big)^{n}$ 
have been found only for a few models 
(the compact massless boson, the critical Ising model and the massless Dirac model)
\cite{cct-09, cct-11, cct-neg, neg-ising,ctt-13-many}.
However, since these formulas have a nonalgebraic form,
the replica limits (\ref{ee-replica-def}) and (\ref{neg-replica-def}) 
for a generic configuration of the two intervals are very challenging
(see Ref.~\cite{DeNo-Coser-extrapolation} for numerical extrapolations).

In this Letter we express the analytic formulas for the moments
$\textrm{Tr} \rho_A^n$ and $\textrm{Tr} \big(\rho_A^{T_2}\big)^{n}$ in algebraic form
for the critical Ising model and the massless Dirac model.
This is a useful step towards the analytic continuations
(\ref{ee-replica-def}) and (\ref{neg-replica-def}).
Furthermore, we relate the resulting expressions 
to the 2D Coulomb gas model on specific lattices.

The two component classical 2D Coulomb gas
is  a neutral mixture of point particles with positive and negative charge $\pm q$.
Their interaction potential, 
repulsive (attractive) for particles with the same (opposite) charges,
is proportional to $q^2 \log (d/a)$, 
where $d$ is the distance between the particles
and $a$ is some length scale \cite{zinn-justin,coulomb-gas-reviews}.
At  inverse temperature $\beta$,
the dimensionless coupling constant in the Boltzmann factor
is $\Gamma = \beta q^2$
and the model exhibits a Kosterlitz-Thouless phase transition at $\Gamma_c = 4$ \cite{KT-paper}.
%
The necessity to introduce a short range repulsion led to studies of the
2D Coulomb gas also on certain lattices \cite{Gaudin-85, Forrester-86, Forrester-Jancovici-sphere}.
When $\Gamma =2$, the model is solvable \cite{Gaudin-85,Gamma2-fermion-equivalence}.
%

We find that the moments $\textrm{Tr} \rho_A^n$ and $\textrm{Tr} \big(\rho_A^{T_2}\big)^n$
for the Ising and Dirac models can be written as the grand canonical partition functions 
of the two-component classical 2D Coulomb gas on certain circular lattices 
at specific values of $\Gamma$, 
which are $\Gamma = 1/2$ for the Ising model and $\Gamma = 1$ for the Dirac model. 
The 2D Coulomb gas model corresponding to the R\'enyi entropies
is very similar to the lattice discretization \cite{Forrester-86}
of the Coulomb gas originally introduced 
to study the simplest Kondo problem \cite{anderson-kondo}.

\newpage

\textit{Entanglement of two disjoint intervals in CFT.---}
In a $1+1$ dimensional CFT on the line and in the ground state,
$A = A_1 \cup A_2$ where $A_1 =(u_1, v_1) $ and $A_2 =(u_2, v_2)$
with $u_1 < v_1 < u_2< v_2$.
The moments $\textrm{Tr} \rho_A^n$ and $\textrm{Tr} \big(\rho_A^{T_2}\big)^n$
contain all the CFT data of the model
(the central charge $c$, the conformal spectrum, and the OPE coefficients)
\cite{EE-2int-initial, cct-09, cct-11, cct-neg}.
They can be obtained as the four-point functions 
$ \textrm{Tr} \rho_A^n = 
\langle \mathcal{T}_n (u_1) \,\overline{\mathcal{T}}_n (v_1)
 \,\mathcal{T}_n (u_2) \,\overline{\mathcal{T}}_n (v_2) \rangle $
 and
  $ \textrm{Tr} \big(\rho_A^{T_2}\big)^n = 
\langle \mathcal{T}_n (u_1) \,\overline{\mathcal{T}}_n (v_1)
 \,\overline{\mathcal{T}}_n (u_2) \,\mathcal{T}_n (v_2) \rangle $
 on the sphere of the twist fields  $\mathcal{T}_n$ 
 and of their conjugate fields $\overline{\mathcal{T}}_n$.
The ordering of the fields  given by  the sequence of the end points
is crucial.
The global conformal invariance on the sphere leads to
\cite{cct-09, cct-11, cct-neg}
\be
\label{R-N-functions-def}
\textrm{Tr} \rho_A^n
=
c_n^2 P_A^{2 \Delta_n }  \mathcal{R}_n(x) \,,
\quad
\textrm{Tr} \big(\rho_A^{T_2}\big)^n
=
c_n^2 P_A^{2 \Delta_n }  \mathcal{N}_n(x)\,,
\ee
where $\Delta_n = \tfrac{c}{12}(n - \tfrac{1}{n})$
is the scaling dimension of the twist fields, 
$c_n$ is a constant, 
$x = \frac{(u_1 - v_1)(u_2 - v_2)}{(u_1 - u_2)(v_1 - v_2)}$ 
is the harmonic ratio of the endpoints of the two intervals 
and
$ P_A =  \tfrac{1}{(v_1 - u_1)(v_2- u_2)(1-x)}$.
We remark that $x\in (0,1)$.
The functions $\mathcal{R}_n(x)$ and $\mathcal{N}_n(x)$ 
originate from the same function $\mathcal{F}_n(z)$ 
with $z \in \mathbb{C}$ as follows 
\be
\label{R-N-def}
\mathcal{R}_n(x) = \mathcal{F}_n(x)\,,
\quad
\mathcal{N}_n(x) = (1-x)^{4\Delta_n} \mathcal{F}_n\bigg(\frac{x}{x-1}\bigg) \,.
\ee

Equivalently, the moments  in Eq.~(\ref{R-N-functions-def}) can be evaluated 
as the partition functions of the CFT 
on the one-parameter family of Riemann surfaces 
defined by the complex curve
\begin{equation}
\label{ZN}
\mathcal{C}
=
\big\{(\lambda,\mu) \in \mathbb{C}^2\,|\;\mu^n = \lambda \, (\lambda-1) \, (\lambda-z)^{n-1}\big\}\,,
\end{equation}
where $z \in \mathbb{C} \setminus\{ 0,1\}$. 
These Riemann surfaces have $\mathbb{Z}_n$ symmetry and genus $g=n-1$. 
To determine $\textrm{Tr} \rho_A^n$ and $\textrm{Tr} \big(\rho_A^{T_2}\big)^n$,
we have to consider the Riemann surfaces corresponding to
$z=x$ and $z=\tfrac{x}{x-1}$, respectively.

The  period matrix   $\tau_n(z)$  of the  curve $\mathcal{C}$ with respect to a given  canonical homology basis   takes the form \cite{cct-09, cct-neg} 
\be
\label{taun}
\tau_n(z)_{i,j} 
 \,= \,
 \frac{2}{n} \sum_{k\,=\,1}^{n-1} \sin(\pi k/n) \,\tau_{k/n}(z)  \cos[2\pi k (i-j)/n]\,,
\ee
where $\tau_{p}(z) =  \textrm{i}  \tfrac{_2F_1(p , 1-p ; 1 ; 1-z)}{_2F_1(p , 1-p ; 1 ; z)} $,
with $\, _2F_1(a , b ; c ; z)$ being the hypergeometric function. 
The Riemann theta function $\Theta[\boldsymbol{e}] (\tau_n(z))$  
with characteristic $\boldsymbol{e}^{\textrm{t}} = (\boldsymbol{\delta}^{\textrm{t}} , \boldsymbol{\varepsilon}^{\textrm{t}} )\in\mathbb{C}^{2(n-1)}$
is defined as \cite{Fay-book}
\be
\label{theta-def}
\Theta[\boldsymbol{e}] (\tau_n(z)) = \!\!
\sum_{\boldsymbol{m} \in\mathbb{Z}^{n-1}} \!\!
\textrm{e}^{ 
\textrm{i} \pi (\boldsymbol{m} + \boldsymbol{\varepsilon})^{\textrm{t}} \,\tau_n(z)  \, (\boldsymbol{m} + \boldsymbol{\varepsilon})
 + 2 \pi \textrm{i} ( \boldsymbol{m} + \boldsymbol{\varepsilon})^{\textrm{t}}  \boldsymbol{\delta}
 }.
\ee

We focus on the CFTs given by the Ising model and the Dirac model,
whose central charges are $c=1/2$ and $c=1$, respectively. 
By employing some results about CFTs of orbifolds 
and on higher genus Riemann surfaces \cite{cft-Riemann-surfaces},
it was found that $\mathcal{F}_n(z)$ in (\ref{R-N-def}) is
\cite{cct-09, cct-11, cct-neg, neg-ising,ctt-13-many}
\be
\label{Fn-ising-dirac}
\mathcal{F}_n(z)
=
\frac{1}{2^{n-1}}
\sum_{\boldsymbol{e}}
\left| 
\frac{\Theta[\boldsymbol{e}] (\tau_n(z))}{
 \Theta[\boldsymbol{0}] ( \tau_n(z))}
 \right|^{2\gamma}\,,
\ee
where $\gamma =1/2$ for the Ising model and $\gamma =1$ for the Dirac model,
and the sum runs over all
half-integer characteristics $\boldsymbol{e}^{\textrm{t}} = (\boldsymbol{\delta}^{\textrm{t}} , \boldsymbol{\varepsilon}^{\textrm{t}} )$,  namely $\delta_j, \varepsilon_j\in\{0,\frac{1}{2}\}$.
%
%
The sum in Eq.~(\ref{Fn-ising-dirac}) contains $\tfrac{1}{2} \binom{2n}{n}$ terms because 
the Riemann theta function (\ref{theta-def}) is nonvanishing only for the
$\tfrac{1}{2} \binom{2n}{n}$ nonsingular even  half-integer characteristics \cite{Fay-book}.

Since the ground state is pure, $S_A^{(n)} = S_B^{(n)}$,
which leads to $\mathcal{R}_n(1-x) = \mathcal{R}_n(x)$.
This relation is found also from the modular invariance of the partition function
\cite{cct-09,cct-11}.
The fermionic model obtained 
by taking only the term with $\boldsymbol{e} = \boldsymbol{0}$ 
in (\ref{Fn-ising-dirac}), which is not modular invariant, 
has also been explored \cite{non-inv-dirac-fermion}.

The CFT formulas for $\textrm{Tr} \rho_A^n$ and $\textrm{Tr} \big(\rho_A^{T_2}\big)^n$
resulting  from Eqs.~(\ref{R-N-functions-def}), (\ref{R-N-def}) and (\ref{Fn-ising-dirac})
with either $\gamma=1/2$ or $\gamma=1$
have been checked through numerical analyses
for the critical Ising chain 
and for the XX chain at the critical point, respectively 
\cite{spin-chain-checks,Coser:2015dvp}.
Finding the analytic continuations required by
the replica limits (\ref{ee-replica-def}) and (\ref{neg-replica-def}) 
with the expression (\ref{Fn-ising-dirac}) 
for any $x\in (0,1)$ is a very challenging task
\cite{cct-11,DHoker:2020bcv}.
Numerical extrapolations have been studied \cite{DeNo-Coser-extrapolation}.
\\


\textit{Hyperelliptic covering and Thomae formula.---}
In this section we reduce the expression of $\mathcal{F}_n$ in  Eq.~\eqref{Fn-ising-dirac}  to an algebraic expression in $z$
by applying the Thomae formula for hyperelliptic curves.
A  hyperelliptic curve of genus $n-1$ has the form $\nu^2=P_{n}(w)$, 
where $P_{n}(w)$ is a polynomial of degree $2n$ and $\nu,w\in\mathbb{C}$.
In 1870 Thomae \cite{Thomae}
showed that,  for hyperelliptic curves,  
the Riemann theta function  \eqref{theta-def} with  $e_j\in\{0,\frac{1}{2}\}$, $j=1,\dots, 2n-2$,
 is proportional to  an algebraic expression in terms of the zeros of the polynomial $P_{n}(w)$.

We  first  recognize  that, 
under the change of  coordinates given by
  $w=\frac{\mu}{\lambda-z}$  and $\nu= (\lambda^2-2\lambda z+z)/(z-\lambda)$,
the curve \eqref{ZN} becomes hyperelliptic \cite{Enolski-Grava-03}
\begin{equation}
\label{hyperell}
{\mathcal C}_{\textrm{\tiny hyp}} 
=
\big\{ (w,\nu)\in\mathbb{C}^2\,|\,
 \nu^2=w^{2n}+2(1-2z)w^n+1 \big\} \, .
\end{equation}

The  zeros of the polynomial   $w^{2n}+2(1-2z)w^n+1$ are
\begin{equation}
 \label{roots-xi-n}
w^{\pm}_{j}(z)=\xi^\pm_n (z)\, \textrm{e}^{\frac{2\pi \textrm{i} j}{n}}\,,
\quad 
\xi^\pm_n (z)=\big( \sqrt{z} \pm \sqrt{z-1}\,\big)^{2/n}\,,
\end{equation}
where $1\leqslant j \leqslant n$. 
We remark that $\xi^+_n (z)\,\xi^-_n (z)=1$.
The points $w^\pm_j$  define the  
circular lattice $\mathcal{I}^{+}_0 \cup \mathcal{I}^{-}_0$  
in the plane, where  
\be
\label{I-0-pm-lattices}
\mathcal{I}^{\pm}_0
=
\big\{\,
w^{\pm}_{j}  = \textrm{e}^{2\pi \textrm{i} j/n} \, \xi^\pm_n(z) \,; 
\, 1\leqslant j \leqslant n \,
\big\}\,.
\ee
\noindent

\begin{figure}[t]
\vspace{-.2cm}
\begin{center}
\includegraphics[width=.48\textwidth]{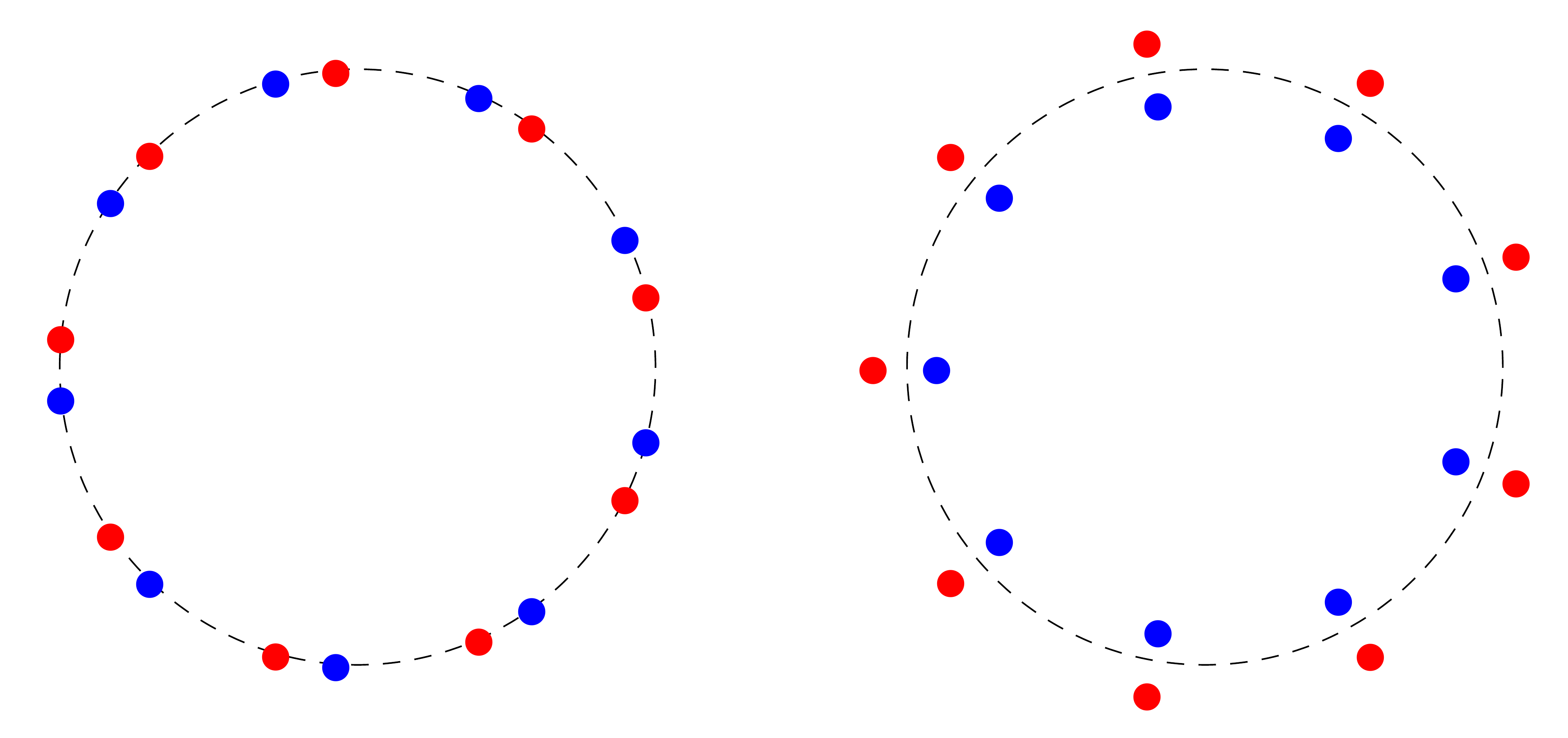}
\end{center}
\vspace{-.6cm}
\caption{
Lattices $\mathcal{I}_0^{+}$ (red circles) and $\mathcal{I}_0^{-}$ (blue circles) in (\ref{I-0-pm-lattices})
for $\textrm{Tr} \rho_A^{n}$ (left) and $\textrm{Tr} \big(\rho_A^{T_2}\big)^{n}$ (right),
when $n=9$ and $x=0.2$ 
(in both panels the unit circle is  indicated by the dashed line).
}  
\label{figure-lattice}
\end{figure}

For the R\'enyi entropies,  $z=x\in(0,1)$,
hence $|\xi^+_n(x)|=1$ and the points $w^{\pm}_{j}(x)$ lie on the unit circle.
In this case the sublattices $\mathcal{I}^{+}_0$ and $\mathcal{I}^{-}_0$ are interwoven
along the unit circle
and the harmonic ratio $x$ parameterizes their angular separation
(see Fig.\,\ref{figure-lattice}, left panel).
Moreover, we have that $\xi^+_n(1-x)= \textrm{e}^{\textrm{i}\pi /n} \xi^-_n(x)$.

Considering the moments $\textrm{Tr} \big(\rho_A^{T_2}\big)^{n}$,
by setting $z =\tfrac{x}{x-1}$  in Eq.~(\ref{roots-xi-n}) one finds
$ \xi^+_n\big( \tfrac{x}{x-1} \big) 
= \big( \tfrac{\sqrt{x} + 1}{\sqrt{1-x}} \big)^{2/n}  \textrm{e}^{\textrm{i}\pi/n}$;
hence in this case the harmonic ratio $x$ parameterizes the radial separation between 
$\mathcal{I}^{+}_0$ and $\mathcal{I}^{-}_0$
(see Fig.\,\ref{figure-lattice}, right panel).

An appropriate choice of the homology basis of the  Riemann surface of the curve 
${\mathcal C}_{\textrm{\tiny hyp}}$ provides
the period matrix of the surface in the form \eqref{taun} \cite{gkt-bis,Enolski-Grava-03};
hence Eq.~\eqref{Fn-ising-dirac} can also be associated to the hyperelliptic curve \eqref{hyperell}.

For hyperelliptic curves, 
the nonsingular even half-integer characteristics  $\boldsymbol{e}$ are in one-to-one correspondence with   
the elements of the set ${\mathcal P}_n$ of  partitions  of the    branch  points (\ref{roots-xi-n})   into   two 
 subsets of  cardinality  $n$ \cite{Fay-book}. 
For our choice of the homology basis of ${\mathcal C}_{\textrm{\tiny hyp}}$ and the base point of the Abel map, 
    the characteristic  $\boldsymbol{e}=\boldsymbol{0}$ corresponds  to  the partition $\{ \mathcal{I}^{+}_0, \mathcal{I}^{-}_0\}$, with $\mathcal{I}^{\pm}_0$ defined in Eq.~\eqref{I-0-pm-lattices}.
A correspondence between the characteristics $\boldsymbol{e}$
and the terms occurring in
$\textrm{Tr} \rho_A^n$ 
and $\textrm{Tr} \big(\rho_A^{T_2}\big)^n$
for the critical Ising chain and the XX chain at the critical point
has been also discussed \cite{Coser:2015dvp}.

The  Thomae formula \cite{Thomae} expresses  the Riemann theta functions 
with nonsingular half-integer characteristics in Eq.~\eqref{theta-def}
as a function of  the branch  points of  the hyperelliptic curve \eqref{hyperell}.  
Generalizations appeared in  Refs.~\cite{EG, Naka,Bern}.
The crucial observation that the curve \eqref{ZN} has the hyperelliptic cover \eqref{hyperell} is essential to apply the original Thomae formula.
 For example, 
 $\big| \Theta[\boldsymbol{0}] (\tau_n(z)) \big|^4
=C \,\big| \Delta(\mathcal{I}^{+}_0)\, \Delta(\mathcal{I}^{-}_0)\big|$,
where $C$ is a constant (irrelevant for our purpose) and 
\be
\Delta(\mathcal{I}^{+}_0)=\prod_{i<j}(w_i^+-w_j^+)
\nonumber
\ee
is the Vandermonde determinant of the points in $\mathcal{I}^{+}_0$, 
and similarly for $\Delta(\mathcal{I}^{-}_0)$.  
For any partition  $\{\mathcal{I}^{+}, \mathcal{I}^{-}\}\in{\mathcal P}_n$,
we denote by $\Delta(\mathcal{I}^{\pm})$ the Vandermonde determinant 
of the points contained in $\mathcal{I}^{\pm}$ 
and by $\boldsymbol{e}$ the 
even half-integer characteristic  associated to the partition  
$\{\mathcal{I}^{+}, \mathcal{I}^{-} \}$. 
Then  the Thomae formula gives 
\begin{equation}
\label{thetan}
\big| \Theta[\boldsymbol{e}] (\tau_n(z)) \big|^4 
=
C\,\big|\Delta(\mathcal{I}^{+}) \, \Delta(\mathcal{I}^{-}) \big|\,,
\end{equation}
where the constant $C$ is independent from $\boldsymbol{e}$.
The Thomae formulas \eqref{thetan} allow to write Eq.~\eqref{Fn-ising-dirac} as
\be
\label{F_n thomae}
\mathcal{F}_n(z) 
=
\frac{1}{2^{n-1}}
\sum_{\{\mathcal{I}^{+}, \mathcal{I}^{-} \}\in{\mathcal P}_n} \!
\left|\, 
\frac{\Delta(\mathcal{I}^{+}) \, \Delta(\mathcal{I}^{-})}
{\Delta(\mathcal{I}^{+}_0) \, \Delta(\mathcal{I}^{-}_0)}
\,\right|^{\gamma/2}.
\ee
Notice that $|\Delta(\mathcal{I}^{+}_0)\,\Delta(\mathcal{I}^{-}_0)|=n^n$. Now $\mathcal{F}_n(z) $ has an algebraic dependence in $z$.
This significantly simplifies both the numerical evaluation of the moments for large $n$ 
and the analysis of their short length expansions \cite{cct-11}. 
\\


\textit{A 2D Coulomb gas in circular lattices.---}
Denote by  $\{\mathcal{I}^{+}_r, \mathcal{I}^{-}_r \}\in{\mathcal P}_n$  the partition obtained from  $\{\mathcal{I}^{+}_0, \mathcal{I}^{-}_0 \}$ by exchanging
$r$ elements between $\mathcal{I}^{+}_0$ and $\mathcal{I}^{-}_0$.
For each $r< n/2$ there is an equal partition of the form  $\{\mathcal{I}^{+}_{n-r}, \mathcal{I}^{-}_{n-r} \}$.
%
Given $0 \leqslant r \leqslant n$,  the number of partitions of the form 
 $\{\mathcal{I}^{+}_r, \mathcal{I}^{-}_r \}$ is $d_{n,r}\binom{n}{r}^2$   where $d_{n,r} = 1-\tfrac{1}{2} \, \delta_{r,n/2}$,  thus  the cardinality of $\mathcal{P}_n$  is
$\tfrac{1}{2} \binom{2n}{n}= \tfrac{1}{2} \sum_{r=0}^{n} \binom{n}{r}^2 = \sum_{r=0}^{\lfloor n/2 \rfloor} d_{n,r}\binom{n}{r}^2 $.
Let us introduce
\be
\label{Fn r-fixed-def}
\mathcal{F}_{n,r}(z) 
\,=\,
\frac{1}{d_{n,r}\,n^{n \gamma/2}}
\sum_{\{\mathcal{I}^{+}_r, \mathcal{I}^{-}_r \}\in{\mathcal P}_n} \!\!\!\!
\big|\Delta(\mathcal{I}^{+}_r) \, \Delta(\mathcal{I}^{-}_r) \big|^{\gamma/2}.
\ee
Since $\mathcal{F}_{n,r}(z) =\mathcal{F}_{n,n-r}(z)$, 
the sum  \eqref{F_n thomae} becomes
\be
\label{F_n thomae-r-sum}
\mathcal{F}_n(z) =
\frac{1}{2^n}
\sum_{r = 0}^{n}  \mathcal{F}_{n,r}(z) 
=
\frac{1}{2^{n-1}}\,
\bigg\{ 1 + \!
\sum_{r = 1}^{\lfloor n/2 \rfloor} \!
d_{n,r} \, \mathcal{F}_{n,r}(z) 
\bigg\}\,,
\ee
where 
we used  
$\mathcal{F}_{n,0}(z) =1$.

After some nontrivial algebra \cite{gkt-bis}, 
we find that Eq.~(\ref{Fn r-fixed-def}) can be written as
\be
\label{2D-coulomb-r-form}
\mathcal{F}_{n,r}(z) 
\,=\,
\left| \frac{\zeta(z)}{n} \right|^{\gamma r}
\sum_{\boldsymbol{i}(r)}\,
\sum_{\boldsymbol{j}(r)}
\big| D_{n,r}(z; \boldsymbol{i} ,\boldsymbol{j}) \big|^\gamma\,,
\ee
where $\zeta(z) =4 \sqrt{z(1-z)}= -\textrm{i} \big( \xi_n^+(z)^n-\xi_n^-(z)^n \big)$ 
and
\be
\label{coulomb-gas-D-term}
D_{n,r}(z; \boldsymbol{i} ,\boldsymbol{j})
\,=\,
\frac{
\prod\limits_{1\leqslant a<b\leqslant r} \!
\big(  w^{+}_{i_b}  - w^{+}_{i_a} \big) \; \big(  w^{-}_{j_b}  - w^{-}_{j_a} \big) 
}{
\prod\limits_{a,b=1}^r \! \big( w^{+}_{i_a}  - w^{-}_{j_b}  \big)} \,.
\ee
In (\ref{2D-coulomb-r-form}),
the sum over $\boldsymbol{i}(r)$ is defined as 
the multiple sum over the $r$-dimensional vectors $\boldsymbol{i}$
made by integers $i_a$ such that $1 \leqslant i_1 <  i_2< \dots < i_r \leqslant n$,
and similarly for the sum over $\boldsymbol{j}(r)$.
These multiple sums can be taken over $1 \leqslant i_a  \leqslant n$
and $1 \leqslant j_b  \leqslant n$, introducing also a multiplicative factor $1/(r!)^2$.
The Cauchy's double alternant formula allows to write
Eq.~(\ref{coulomb-gas-D-term}) as a single determinant.
Notice that the numerator of Eq.~(\ref{coulomb-gas-D-term})
is independent of $z$.

From Eqs.~(\ref{F_n thomae-r-sum}), (\ref{2D-coulomb-r-form}) and (\ref{coulomb-gas-D-term})
we recognize that $2^{n} \mathcal{F}_n(z) $ is the grand canonical partition function
of the 2D classical Coulomb gas on a lattice where 
 the positive and negative charges are constrained to occupy the sites of
$\mathcal{I}_0^{+}$ and $\mathcal{I}_0^{-}$, respectively.
Each site can be either empty or occupied by one particle. 
The parameter $\gamma$ is identified with the 
dimensionless coupling constant $\Gamma = \beta q^2$
of the Coulomb gas,
the expression  $|\zeta(z)/n|$ with $\lambda a$, 
where $\lambda$ is the fugacity,
and the integer $0 \leqslant r \leqslant n$ 
with the number of positive charges in $\mathcal{I}_0^{+}$ 
and of negative charges in $\mathcal{I}_0^{-}$.

The Coulomb gas with $\Gamma =1$ on the circular lattice studied in Ref.~\cite{Forrester-86}
is closely related to $\textrm{Tr} \rho_A^n$ at $x=1/2$ for the Dirac model.
%
As for the moments $\textrm{Tr} \big(\rho_A^{T_2}\big)^{n}$, 
it can be insightful to map the corresponding sublattices $\mathcal{I}^{\pm}_0$
on the sphere through the stereographic projection \cite{Forrester-Jancovici-sphere}.

For the R\'enyi entropies, 
$w_j^{\pm}(1-x) = \textrm{e}^{\pi \textrm{i} /n}w_j^{\mp}(x)$.
This implies that, in Eq.~(\ref{2D-coulomb-r-form}), $\mathcal{F}_{n,r}(1-x) = \mathcal{F}_{n,r}(x)$ for any $r$.
Thus, we reobtained the relation $\mathcal{R}_n(1-x) = \mathcal{R}_n(x)$,
as expected from the purity of the ground state and from the modular invariance \cite{cct-11}.

When $r=1$,
the expression (\ref{2D-coulomb-r-form}) can be written as
\be
\label{F-n-1-sum}
\mathcal{F}_{n,1}(z) 
=
n
\sum_{k=1}^{n}
\bigg|
 \frac{(\xi_n^{+})^n - (\xi_n^{-})^n }{n \big(\xi^{+}_n - \eta_n^k \,\xi^{-}_n\big)} 
 \bigg|^{\gamma}\!
 =\!
\sum_{p = 1}^{\lceil n/2 \rceil} \!
 d_{n+1,p} \, \mathsf{F}_{n,1}^{(p)}(z)\,,
\ee
where $ \eta_n= \textrm{e}^{2\pi \textrm{i} /n} $
and $ \mathsf{F}_{n,1}^{(p)}(z)$ is defined as
the sum of two summands with $k=p$ and $k= n-p+1$.
%
We remark that, considering the R\'enyi entropies, we have
$\mathsf{F}_{n,1}^{(p)}(1-x)  = \mathsf{F}_{n,1}^{(p)}(x) $ 
for any $1\leqslant p \leqslant \lceil n/2 \rceil$.

The large $n$ limit, which allows to study
the largest eigenvalues of $\rho_A$ and $\rho_A^{T_2}$,
can be explored through the Coulomb gas in the continuum,
also by employing its equivalence with the sine-Gordon model \cite{zinn-justin,samaj-book}.
For instance,  when $\gamma < 1$,  the leading order of  Eq.~(\ref{F-n-1-sum}) as $n \to \infty$
is $n^{2-\gamma} | \zeta(z)|^\gamma  \tfrac{\Gamma(1-\gamma)}{\Gamma(1-\gamma/2)^2}$,
in agreement with Ref.~\cite{fendley-saleur}.
\\


\textit{Special cases.---}
%
When $n=2$, the tori occurring in 
$\textrm{Tr} \rho_A^n$ 
and $\textrm{Tr} \big(\rho_A^{T_2}\big)^n$
are equivalent because their modular parameters are related 
through a modular transformation \cite{cct-neg}.
In this case, the last expression in Eq.~(\ref{F_n thomae-r-sum}) 
contains only the term (\ref{F-n-1-sum}) specified to $n=2$
and Eq.~(\ref{R-N-def}) becomes
\be
\mathcal{R}_2(x) = \mathcal{N}_2(x) = \frac{1}{2} \, \Big\{ 1+ x^{\gamma/2} + (1-x)^{\gamma/2} \Big\}\,,
\nonumber
\ee
which is invariant under $x \leftrightarrow 1-x$, as expected.

When $n=3$, the genus two Riemann surfaces  
for $\textrm{Tr} \rho_A^n$ and $\textrm{Tr} \big(\rho_A^{T_2}\big)^n$
are not equivalent. 
Also in this case only  (\ref{F-n-1-sum}) 
contributes to 
$\mathcal{F}_3(z)= \tfrac{1}{4}\big[1+\mathsf{F}_{3,1}^{(1)}(z)+\mathsf{F}_{3,1}^{(2)}(z)\big]$ with 
\bea
\mathsf{F}_{3,1}^{(1)}(z)
&=&
\frac{3}{3^\gamma} \Big[\,
\big| \big( \xi_3^{+} - \xi_3^{-} \big) \, \big( \xi_3^{+} - \eta_3^2\, \xi_3^{-} \big) \big|^\gamma
\nonumber
\\
& & \hspace{.7cm}
+\, 
\big| \big( \xi_3^{+} - \eta_3\,\xi_3^{-} \big) \, \big( \xi_3^{+} - \eta_3^2\, \xi_3^{-} \big) \big|^\gamma
\, \Big]
\nonumber
\\
\rule{0pt}{.5cm}
\mathsf{F}_{3,1}^{(2)}(z)
&=&
\frac{3}{3^\gamma} 
\big| \big( \xi_3^{+} - \xi_3^{-} \big) \, \big( \xi_3^{+} - \eta_3\, \xi_3^{-} \big) \big|^\gamma\,.
\nonumber
\eea
%

In the case of $n=4$, the genus three Riemann surfaces  
for $\textrm{Tr} \rho_A^n$ and $\textrm{Tr} \big(\rho_A^{T_2}\big)^n$
are not equivalent too. 
The last expression in Eq.~(\ref{F_n thomae-r-sum})  becomes
\be
\mathcal{F}_4(z) =
\frac{1}{2^3}
\left(1 +\mathcal{F}_{4,1}(z) + \frac{\mathcal{F}_{4,2}(z)}{2}   \,\right)\,,
\nonumber
\ee
where $\mathcal{F}_{4,1}(z)$ is 
(\ref{F-n-1-sum}) specialized to  $n=4$ and 
\be
\label{F-4-2-sum}
\mathcal{F}_{4,2}(z) 
=
\frac{16\,|\zeta(z)|^{\gamma}}{8^{\gamma}}
\left(
2^{\frac{\gamma}{2}}
+
 \frac{\mathsf{F}_{4,2}^{(1)}(z)}{8}
+
\frac{\mathsf{F}_{4,2}^{(2)}(z) + \mathsf{F}_{4,2}^{(3)}(z)}{4}
\right)
\nonumber
\ee
with 
\bea
\mathsf{F}_{4,2}^{(1)}(z)
& = &
\bigg| 
\frac{2\, \zeta(z)}{(\xi_4^{+} - \xi_4^{-})^2 \,(\xi_4^{+} - \eta_4^2\, \xi_4^{-} )^2} 
\bigg|^{\gamma}
\nonumber
\\
\rule{0pt}{.7cm}
& & +\;
\bigg| 
\frac{2\, \zeta(z)}{(\xi_4^{+} - \eta_4\, \xi_4^{-} )^2 \,(\xi_4^{+} - \eta_4^3\, \xi_4^{-} )^2} 
\bigg|^{\gamma}
\nonumber
\eea
and
\bea
\mathsf{F}_{4,2}^{(2)}(z)
& = &
\bigg| 
\frac{\xi_4^{+} -  \xi_4^{-}}{\xi_4^{+} - \eta_4^2\, \xi_4^{-}} 
\bigg|^{\gamma}
\! +\,
\bigg| 
\frac{\xi_4^{+} - \eta_4\, \xi_4^{-}}{\xi_4^{+} - \eta_4^3\, \xi_4^{-}} 
\bigg|^{\gamma}
\nonumber
\\
\rule{0pt}{.8cm}
\mathsf{F}_{4,2}^{(3)}(z)
& = &
\bigg| 
\frac{\xi_4^{+} - \eta_4^2\, \xi_4^{-}}{\xi_4^{+} - \xi_4^{-}} 
\bigg|^{\gamma}
\!+\,
\bigg| 
\frac{\xi_4^{+} - \eta_4^3\, \xi_4^{-}}{\xi_4^{+} - \eta_4\, \xi_4^{-}} 
\bigg|^{\gamma}.
\nonumber
\eea
When $z=x\in (0,1)$,
each $\mathsf{F}_{4,2}^{(p)}(x)$ is $x \leftrightarrow 1-x$ invariant.
\\


\textit{Conclusions.---}
We studied the moments of the reduced density matrix $\textrm{Tr} \rho_A^n$ 
and of its partial transpose $\textrm{Tr} \big(\rho_A^{T_2}\big)^n$ 
when $A$  is the union of two disjoint intervals on the line,
in the CFTs given by the Ising and the Dirac model in their ground state.
We found that the existing results, given by 
Eqs.~(\ref{R-N-functions-def}) and (\ref{Fn-ising-dirac}),
can be expressed through Eq.~(\ref{F_n thomae}), 
where the dependence on the harmonic ratio $x$ is algebraic. 
This significantly simplifies the numerical evaluation of these quantities. 
Furthermore, this leads to
Eqs.~(\ref{F_n thomae-r-sum}), (\ref{2D-coulomb-r-form}) and (\ref{coulomb-gas-D-term}),
establishing a remarkable equivalence between the moments 
$\textrm{Tr} \rho_A^n$ and $\textrm{Tr} \big(\rho_A^{T_2}\big)^n$
and  the grand partition functions of the
classical 2D Coulomb gas in the circular lattices defined by Eq.~(\ref{I-0-pm-lattices})
(see Fig.\,\ref{figure-lattice})
at $\Gamma = 1/2$ for the Ising model
and at $\Gamma = 1$ for the Dirac model.

These results provide a new tool to tackle 
the analytic continuations (\ref{ee-replica-def}) and (\ref{neg-replica-def}),
in order to obtain $S_A$ and $\mathcal{E}$ analytically for $x \in (0,1)$.
We find it worth exploring also the limit of large $n$ \cite{gkt-bis}.
It would be interesting to extend our analysis by considering 
other models (e.g. the massless compact boson \cite{cct-09})
or more complicated configurations 
(e.g., when $A$ is made by $N$ disjoint intervals \cite{ctt-13-many}) 
or other physically relevant situations 
(e.g., when the temperature or the volume of the system are finite \cite{datta-david, cct-neg-temp}).
In particular, understanding whether an equivalence similar to the one 
found in this Letter occurs also when the central charge is large
(considering the system on the line and in its ground state, 
different partition functions on the family of Riemann surfaces (\ref{ZN}) must be studied)
would provide useful insights in the study of entanglement in the AdS/CFT
correspondence \cite{ent-ads}.

\vspace{1.1mm}
 

\noindent
We are grateful to Andrea Coser, Riccardo Fantoni, 
Mihail Mintchev and, in particular, to Giuseppe Mussardo for useful discussions. 
T.G.  acknowledges support from the European Union's H2020 research and innovation program under the Marie Sk\l owdoska--Curie grant No. 778010 {\em  IPaDEGAN}  and the support of INDAM/GNFM.
%


\end{document}